\documentclass[10pt,twocolumn]{article}
\usepackage{graphicx}
\usepackage{textcomp}
\usepackage{amsmath}
\usepackage[hmargin=1.5cm,vmargin=2.0cm]{geometry}
\usepackage[font=small,margin=0.4cm,figurename=FIGURE]{caption}
\usepackage[sort&compress,numbers]{natbib}
\usepackage{abstract}
\usepackage{hyperref}

\hypersetup{
    colorlinks=true,
    linkcolor=black,
    citecolor=black,
    filecolor=black,
    urlcolor=black,
}

\renewcommand{\abstractname}{}    % clear the title
\renewenvironment{abstract}
 {%\small
  \begin{center}
  \bfseries \abstractname\vspace{-1.05cm}\vspace{0pt}
  \end{center}
  \list{}{
    \setlength{\leftmargin}{2cm}%
    \setlength{\rightmargin}{\leftmargin}%
  }%
  \item\relax}
 {\endlist}

\begin{document}

\title{Manipulating waves with LEGO\textsuperscript{\textregistered} bricks:\\ A versatile experimental platform for metamaterial architectures} 

\author{Paolo Celli, Stefano Gonella}
\date{Department of Civil, Environmental, and Geo- Engineering\\University of Minnesota, Minneapolis, MN 55455, USA\\
\vspace{15px} \normalsize{\underline{\bf Published article}: \emph{Appl. Phys. Lett.} {\bf 107}, 081901 (2015);\quad \url{http://dx.doi.org/10.1063/1.4929566}}}

% ABSTRACT
\twocolumn[
\begin{@twocolumnfalse}
\maketitle
\begin{abstract}
In this letter, we discuss a versatile, fully-reconfigurable experimental platform for the investigation of phononic phenomena in metamaterial architectures. The approach revolves around the use of 3D laser vibrometry to reconstruct global and local wavefield features in specimens obtained through simple arrangements of LEGO\textsuperscript{\sffamily\textregistered} bricks on a thin baseplate. The agility by which it is possible to reconfigure the brick patterns into a nearly endless spectrum of topologies makes this an effective approach for rapid experimental proof of concept, as well as a powerful didactic tool, in the arena of phononic crystals and metamaterials engineering. We use our platform to provide a compelling visual illustration of important spatial wave manipulation effects (waveguiding and seismic isolation), and to elucidate fundamental dichotomies between Bragg-based and locally resonant bandgap mechanisms.
\vspace{0.4cm}
\end{abstract}
\end{@twocolumnfalse}
]

% INTRODUCTION
Over the past two decades, acousto-elastic phononic crystals and metamaterials have received increasing attention due to their unique wave manipulation capabilities. Perhaps the most well-known property is the ability to open phononic bandgaps~\cite{Kushwaha_1993,Sigalas_1993,Martinez-Sala_1995,Liu_2000,Martinsson_2003}, i.e., frequency intervals of forbidden wave propagation. Another distinctive feature is the frequency-dependent anisotropy (or directivity) observed in their spatial wave patterns~\cite{Langley_1996,Ruzzene_2003,Gonella_2008} and the ability to feature a negative refractive index~\cite{Yang_2004,Hennion_2013,Zhu_2014}. A number of metamaterial architectures have also been proposed to design acoustic lenses~\cite{Zhang_2009,Li_2009,Spadoni_2010} and attain subwavelength imaging~\cite{Sukhovich_2009,Lemoult_2011,Zhu_2011}.

In recent years, the field of phononics has seen a significant increase in the amount of experimental work~\cite{Hussein_2014}. In the realm of metamaterials design, the need for experiments is pressing on two levels. At the conceptual development stage, it is beneficial to rely on a set of fast and economical, yet accurate experiments to obtain quick \emph{proof of concept}, or to compare and rank multiple candidate configurations. At the later stages of the process, more precise, \emph{ad hoc} experiments are required to fully characterize the selected configuration and to verify its actual implementability as a device. For proof of concept, it is often too costly and time consuming to work with specimens fabricated with advanced manufacturing methods, especially when a large number of configurations must be tested; it may be convenient to work with  \emph{laboratory materials} characterized by low cost, fast manufacturing time and high reconfigurability. These attributes are also crucial in the development of prototypes for didactic demonstrations. 3D-printed materials, for example, display many of these characteristics; however, they are typically affected by unreliable mechanical properties and high geometric variability; moreover, their response is often tainted by high levels of damping that are detrimental to the establishment of noise-free wavefields.

Inspired by these considerations, we here explore a versatile platform for rapid verification of phononic phenomena in metamaterial architectures, based on the reversible assembly of patterns of LEGO\textsuperscript{\sffamily\textregistered} bricks on a thin baseplate. LEGO\textsuperscript{\sffamily\textregistered} components have been widely used, especially for didactic purposes, in a variety of scientific environments, including robotics~\cite{GdG_2011}, biological sciences~\cite{Gothelf_2012,Lind_2014} and physics~\cite{Quercioli_1998,Chen_2014}. In our approach, bricks of different size and shape can be placed at designated locations on a baseplate as to produce a variety of periodic or disordered stub patterns with different length scales. Thanks to the non-permanent brick-plate contacts, it is possible to seamlessly transition between different topologies, effectively switching between radically different tests, in a matter of minutes. Note that the acrylonitrile butadiene styrene (ABS) baseplate, which serves as the wave propagation support, displays relatively low viscosity, which results in low damping and moderate wave attenuation.

For our measurements, we rely on the sensing flexibility of a 3D Scanning Laser Doppler Vibrometer (SLDV), which allows simultaneous in-plane and out-of-plane wavefield reconstruction. Thanks to their sensitivity and frequency bandwidth, and in light of the benefits of non-contact sensing, laser vibrometers have gained increasing popularity for the experimental analysis of phononic structures~\cite{Kokkonen_2007, Oudich_2011, Assouar_2012, Casadei_2012, Celli_JSV_2014, Rupin_2014}. Here, the scanning vibrometer contributes to the sensing side of the problem a dimension of reconfigurability that complements the flexibility that is achieved, at the specimen fabrication level, by using bricks as building blocks.

\begin{figure} [!htb]
\centering
\includegraphics[scale=1.48]{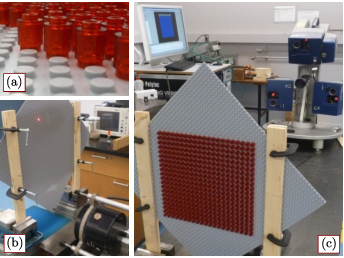}
\caption{Experimental setup. (a) Detail of the baseplate with cylindrical bricks. (b) Scanned surface (backside of the baseplate) and shaker. (c) Front view of the specimen with the three scanning heads of the 3D-SLDV in the background.}
\label{Fig1}
\end{figure}

% THE EXPERIMENTAL SETUP
%\section{Experimental setup} \label{sec:exp}
Throughout this letter, all the considered topologies are obtained as assemblies of cylindrical bricks on a baseplate~\cite{Pennec_2008,Wu_2008}. The baseplate features periodically-distributed \emph{studs}---small cylindrical protuberances where the bricks can be anchored. The arrangement of studs allows for several periodic configurations with various lattice constants. Note that the brick-stud contact only relies on friction: no use of glue (which would undermine agile specimen reconfiguration) is required. A detail of bricks and baseplate is shown in Fig.~\ref{Fig1}a.
In Fig.~\ref{Fig1}b we show the scanned surface (the backside of the plate), coated with a thin layer of reflective paint to increase the quality of the optical signal. The excitation is imparted using a Bruel \& Kjaer Type $4809$ shaker (also shown in Fig.~\ref{Fig1}b), with a $0\mbox{--}20\,\mathrm{kHz}$ bandwidth. Sensing is performed by scanning the baseplate with a Polytec PSV-400-3D Scanning Laser Doppler Vibrometer (Fig.~\ref{Fig1}c). To probe the response over broad frequency ranges, we excite the specimen with a pseudorandom waveform featuring a flat power spectrum over the frequency range of interest. To improve the signal-to-noise ratio, measurements at each scanned location are repeated $20$ times and averaged. A low-pass filter is used to cut-off all spurious features in the signal above $20\,\mathrm{kHz}$~\cite{suppl}.

% BANDGAP DUALITY
%\section{The bandgap duality} \label{sec:bg}
Our first objective is to use our combined fabrication and sensing platform to visually elucidate the phononic bandgap phenomenon. Phononic bandgaps are frequency intervals in which wave propagation is forbidden. The generation of bandgaps can be traced back to two main mechanisms: Bragg scattering and locally resonant effects. Bragg bandgaps are due to the destructive interference between cascades of waves scattered at the internal interfaces in the material. They require periodicity in the structure and manifest when the wavelength of excitation approaches the characteristic size of the unit cell. Locally resonant bandgaps are a byproduct of energy-localization mechanisms; they take place when the frequency of excitation approaches the resonance frequencies of some ``microstructural'' elements~\cite{Martinsson_2003}. They are \emph{subwavelength} in nature, i.e. they arise when the wavelengths are about one order of magnitude larger than the unit cell size. In the following, we leverage the versatility of our experimental setup to illustrate the duality between these mechanisms.

\begin{figure*} [!htb]
\centering
\includegraphics[scale=1.48]{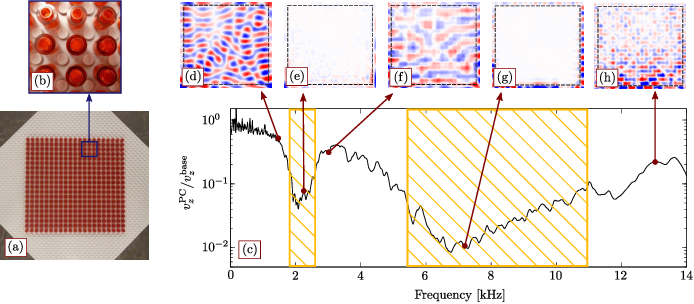}
\caption{Bandgap analysis for a dense periodic brick topology. (a,b) Specimen with detail of the brick arrangement. (c) Frequency response function (average out-of-plane velocity normalized by the average value measured along the bottom edge). The shaded areas highlight the bandgap regions. (d-h) Laser-acquired wavefields sampled at five reference frequencies; the dashed lines delimit the region occupied by the bricks. (d) $1.5\,\mathrm{kHz}$. (e) $2.25\,\mathrm{kHz}$. (f) $3\,\mathrm{kHz}$. (g) $7.25\,\mathrm{kHz}$. (h) $13\,\mathrm{kHz}$.}
\label{Fig2}
\end{figure*}

Our first specimen comprises $420$ bricks arranged as shown in Figs.~\ref{Fig2}a and b.
For convenience, we label this topology ``dense''. Fig.~\ref{Fig2}c shows the frequency response function (FRF) calculated from surface measurements using the vibrometer in the $0\mbox{--}14\,\mathrm{kHz}$ range; the average out-of-plane velocity $v^{\mathrm{PC}}_z$ measured at the scan points inside the brick-filled area (the actual phononic crystal zone) is divided by the average velocity $v^{\mathrm{base}}_z$ measured along the bottom edge of the scanned surface. We observe two bandgaps (shaded areas): a narrow band approximately in the $1.8\mbox{--}2.6\,\mathrm{kHz}$ range, and a wider one approximately in the $5.5\mbox{--}11\,\mathrm{kHz}$ range. Snapshots of the steady-state response at selected frequencies are shown in Figs.~\ref{Fig2}d$\mbox{--}$h. In all plots, the velocity ranges from $-0.4\,v^{\mathrm{base}}_{z,\,\mathrm{max}}$ to $0.4\,v^{\mathrm{base}}_{z,\,\mathrm{max}}$, where $v^{\mathrm{base}}_{z,\,\mathrm{max}}$ is the maximum amplitude recorded along the bottom edge of the scanned region at each frequency. The wavefield response provides visual evidence of both bandgaps: Figs.~\ref{Fig2}d, f, h, corresponding to propagation regions, feature appreciable oscillations over the entire scanned area; Figs.~\ref{Fig2}e and g, corresponding to attenuation zones, feature nearly no motion in the crystal region (bordered by the dashed line). A comparison with the FRF of a pristine baseplate, confirming the fact that the bandgaps are due to the presence of the brick arrangement, is given in the supplemental material (SM)~\cite{suppl}.

\begin{figure} [!htb]
\centering
\includegraphics[scale=1.48]{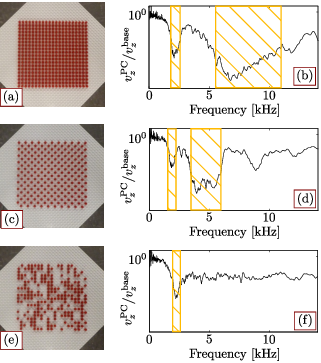}
\caption{Effects of scale coarsening and periodicity relaxation. (a) Dense periodic topology ($420$ bricks). (b) FRF for the topology in (a). (c) Coarse periodic topology ($210$ bricks). (d) FRF for the topology in (c). (e) Random assembly of $210$ bricks. (f) FRF for the topology in (e).}
\label{Fig3}
\end{figure}

To investigate the dual nature of the observed bandgaps, we test their robustness against changes in topology. Here we invoke the notion that Bragg bandgaps are sensitive to changes in lattice spacing and to relaxation of the periodicity, while locally resonant bandgaps only depend upon the availability of resonating elements. In Fig.~\ref{Fig3}, the reference dense topology (Figs.~\ref{Fig3}a and b) is compared to two other topologies that are assembled via rapid reconfiguration of the brick pattern.
The configuration in Fig.~\ref{Fig3}c is obtained by downsampling (by half) the original topology while retaining a periodic arrangement. The corresponding FRF in Fig.~\ref{Fig3}d still displays two main bandgaps: the gap in the neighborhood of $2.2\,\mathrm{kHz}$ is preserved, albeit narrower than in the dense topology; the larger gap has shifted towards lower frequencies ($3.5\mbox{--}6\,\mathrm{kHz}$ range, approximately). Shifting towards lower frequencies (longer wavelengths) in response to increases in lattice constant can be seen as the signature of Bragg scattering, which indeed requires compatibility between unit cell size and wavelength. Fig.~\ref{Fig3}e, in contrast, is obtained by rearranging the $210$ bricks into a (MATLAB-generated) disordered pattern. The FRF in Fig.~\ref{Fig3}f highlights the survival of the $2.2\,\mathrm{kHz}$ bandgap while the higher gap has vanished.  The persistence of the low frequency gap across all topologies highlights its periodicity-independence and hints at underlying locally resonant mechanisms---an aspect recently investigated by Rupin et al.~\cite{Rupin_2014}. The intensity of the resonant mechanisms depends on the number of resonators that can contribute to energy trapping. This confirms the reduction in the first bandgap width in going from the case of Fig.~\ref{Fig3}a to those of Figs.~\ref{Fig3}c, e. In SM~\cite{suppl} we also discuss the influence of the resonators' height on both locally resonant and Bragg bandgaps.

\begin{figure} [!htb]
\centering
\includegraphics[scale=1.48]{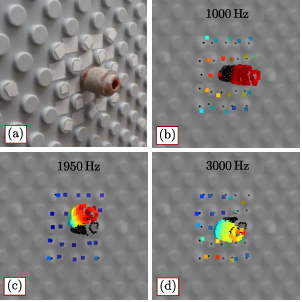}
\caption{Experimental reconstruction of the motion of a cylindrical brick at three frequencies in the neighborhood of the locally resonant bandgap. (a) Detail of the brick. (b) Motion at $1000\,\mathrm{Hz}$, before the presumed bandgap region. (b) Motion at $1950\,\mathrm{Hz}$, near the onset of the bandgap. (d) Motion at $3000\,\mathrm{Hz}$, above the bandgap.}
\label{Fig4}
\end{figure}

To substantiate our hypothesis regarding the locally resonant nature of the $2.2\,\mathrm{kHz}$ bandgap, we attempt to reconstruct experimentally the vibrational behavior of a single brick (mounted at the center of the baseplate) at different frequencies falling inside and outside the bandgap deemed to be locally resonant; a detail of the brick, coated in a thin layer of reflective paint, is shown in Fig.~\ref{Fig4}a.
For this task, we define a dense cylindrical scan grid on the surface of the brick, as well as a coarser rectangular grid on the region of the plate immediately surrounding the brick. The color given to the scanned points is proportional to the RMS of the three measured velocity components; at each frequency, the velocity is normalized by the maximum velocity recorded over the entire scanned region at that frequency, to highlight the relative motion between brick and baseplate. The black dots mark the original position of the scan points. Fig.~\ref{Fig4}b shows the brick-and-plate motion at $1000\,\mathrm{Hz}$ (before the bandgap). We can see that the brick moves in the out-of-plane direction in phase with the plate substrate, which undergoes significant deformation. Fig.~\ref{Fig4}c pinpoints a frequency near the onset of the bandgap, $1950\,\mathrm{Hz}$: the brick undergoes large tilting motion, while the points on the plate are approximately still. Fig.~\ref{Fig4}d represents a post-gap frequency ($3000\,\mathrm{Hz}$) where, once again, we observe significant motion of both the plate and the brick. We conclude that we only observe large relative motion between stub and substrate at the onset of the bandgap: this provides evidence of energy trapping and highlights the locally resonant nature of the bandgap.
%

% PHONONIC PHENOMENA
%\section{waveguiding and shape highlighting} \label{sec:eff}
We can further exploit the reconfigurability of our specimens to illustrate (using the same brick set) a variety of spatial manipulation phenomena, such as waveguiding and seismic isolation, at different wavelengths.
\begin{figure} [htb]
\centering
\includegraphics[scale=1.48]{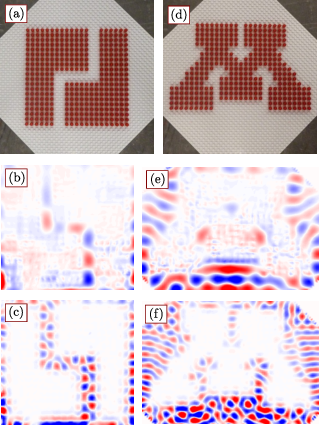}
\caption{Experimental evidence of waveguiding and isolation effects. (a) Snake-like waveguide. (b) Waveguide response at $2512.5\,\mathrm{Hz}$. (c) Waveguide response at $7000\,\mathrm{Hz}$. (d) Metamaterial realization of the Minnesota M logo. (e) Image of the M at $2362.5\,\mathrm{Hz}$. (f) Image of the M at $7000\,\mathrm{Hz}$.}
\label{Fig5}
\end{figure}
The first topology, shown in Fig.~\ref{Fig5}a, is obtained by assembling a dense square phononic crystal and by introducing a two-unit-cell wide snake-like defect path. Defect paths are known to act as waveguides when the frequency of excitation falls inside a bandgap for the crystal. Consistently with the bandgap duality discussed above, waveguiding is attainable for wavelengths that are comparable~\cite{Torres_1999,Jensen_2003,Khelif_2004,Pennec_2009} or larger~\cite{Assouar_2012,Lemoult_2013} than the unit cell. In Fig.~\ref{Fig5}b we obtain long-wavelength waveguiding (at $2512.5\,\mathrm{Hz}$), while in Fig.~\ref{Fig5}c we report waveguiding at shorter wavelengths (at $7000\,\mathrm{Hz}$, inside the Bragg gap).

We then proceed to rearrange the bricks into the topology shown in Fig.~\ref{Fig5}d, which implements a phononic crystal realization of the University of Minnesota M logo. The wavefields obtained at $2362.5\,\mathrm{Hz}$ and $7000\,\mathrm{Hz}$ are shown in Fig.~\ref{Fig5}e and Fig.~\ref{Fig5}f, respectively. We can see how the energy is predominantly confined outside the crystal region; this effect is especially pronounced in the Bragg bandgap, where the response decays inside the M contour within $1\mbox{--}2$ cell layers. This configuration, in conjunction with the ability to scan both interior and exterior of the crystal, provides an eloquent lab demonstration of the potentials of PCs for vibration and seismic isolation~\cite{Meseguer_1999,Brule_2014}. In SM~\cite{suppl}, we discuss two additional crystal topologies to emphasize the breadth of effects that can be verified using the proposed platform.

% CONCLUSIONS
%\section{Conclusions} \label{sec:c}
As a summary, in this letter we have illustrated how, through a simple assembly of LEGO\textsuperscript{\sffamily\textregistered} bricks, we can generate metamaterial architectures with locally resonant and Bragg-based bandgaps. The reconfigurability of the brick patterns, together with the availability of bricks of virtually any size and shape, can be leveraged to assemble and test a plethora of metamaterial architectures, thus serving as a versatile experimental platform for \emph{proof of concept} lab tests. We believe that the intuitive and tangible nature of toy-based specimens, along with the visualization capabilities of laser sensing, can provide a dimension of intuitiveness to the field of experimental phononics and act as perfect platform for far-reaching teaching and outreach activities in the realm of wave mechanics and metamaterials engineering.

% ACKNOWLEDGEMENTS
%\section*{Acknowledgements}
The inspiration for this work sprouted from ideas originated during the organization of outreach activities within a project sponsored by the National Science Foundation (grant CMMI-1266089). We are particularly indebted to Nathan Bausman and Davide Cardella for their insight and assistance with the experiment setup. We also wish to thank Jeff Druce and R. Ganesh for their valuable input.

%\begin{acknowledgments}
%\end{acknowledgments}

\begin{center}
\noindent\rule{3cm}{0.4pt}
\end{center}

\renewcommand\refname{\vskip -1.3cm}
\bibliographystyle{unsrt}
%{\small\bibliography{myrefs.bib}}

{\small
\begin{thebibliography}{39}

\bibitem{Kushwaha_1993}
M.~S. Kushwaha, P.~Halevi, L.~Dobrzynski, and B.~Djafari-Rouhani.
\newblock Acoustic band structure of periodic elastic composites.
\newblock {\em Phys. Rev. Lett.}, 71:2022--2025, 1993.

\bibitem{Sigalas_1993}
M.~Sigalas and E.~N. Economou.
\newblock Band structure of elastic waves in two dimensional systems.
\newblock {\em Solid State Commun.}, 86(3):141--143, 1993.

\bibitem{Martinez-Sala_1995}
R.~Martinez-Sala, J.~Sancho, and J.~V. Sanchez.
\newblock {Sound-attenuation by sculpture}.
\newblock {\em Nature}, 378:241, 1995.

\bibitem{Liu_2000}
Z.~Liu, X.~Zhang, Y.~Mao, Y.~Y. Zhu, Z.~Yang, C.~T. Chan, and P.~Sheng.
\newblock Locally resonant sonic materials.
\newblock {\em Science}, 289(5485):1734--1736, 2000.

\bibitem{Martinsson_2003}
P.~G. Martinsson and A.~B. Movchan.
\newblock Vibrations of lattice structures and phononic band gaps.
\newblock {\em Q. J. Mech. Appl. Math.}, 56(1):45--64, 2003.

\bibitem{Langley_1996}
R.~S. Langley.
\newblock The response of two-dimensional periodic structures to point harmonic
  forcing.
\newblock {\em J. Sound Vib.}, 197(4):447--469, 1996.

\bibitem{Ruzzene_2003}
M.~Ruzzene, F.~Scarpa, and F.~Soranna.
\newblock Wave beaming effects in two-dimensional cellular structures.
\newblock {\em Smart Mater. Struct.}, 12(3):363, 2003.

\bibitem{Gonella_2008}
S.~Gonella and M.~Ruzzene.
\newblock Analysis of in-plane wave propagation in hexagonal and re-entrant
  lattices.
\newblock {\em J. Sound Vib.}, 312(1–2):125--139, 2008.

\bibitem{Yang_2004}
S.~Yang, J.~Page, Z.~Liu, M.~Cowan, C.~Chan, and P.~Sheng.
\newblock Focusing of sound in a 3d phononic crystal.
\newblock {\em Phys. Rev. Lett.}, 93:024301, 2004.

\bibitem{Hennion_2013}
A.-C. Hladky-Hennion, J.~O. Vasseur, G.~Haw, C.~Croënne, L.~Haumesser, and
  A.~N. Norris.
\newblock Negative refraction of acoustic waves using a foam-like metallic
  structure.
\newblock {\em Appl. Phys. Lett.}, 102(14):144103, 2013.

\bibitem{Zhu_2014}
R.~Zhu, X.~N. Liu, G.~K. Hu, C.~T. Sun, and G.~L. Huang.
\newblock Negative refraction of elastic waves at the deep-subwavelength scale
  in a single-phase metamaterial.
\newblock {\em Nat. Commun.}, 5:5510, 2014.

\bibitem{Zhang_2009}
S.~Zhang, L.~Yin, and N.~Fang.
\newblock Focusing ultrasound with an acoustic metamaterial network.
\newblock {\em Phys. Rev. Lett.}, 102:194301, 2009.

\bibitem{Li_2009}
J.~Li, L.~Fok, X.~Yin, G.~Bartal, and X.~Zhang.
\newblock Experimental demonstration of an acoustic magnifying hyperlens.
\newblock {\em Nat. Mater.}, 8(12):931--934, 2009.

\bibitem{Spadoni_2010}
A.~Spadoni and C.~Daraio.
\newblock Generation and control of sound bullets with a nonlinear acoustic
  lens.
\newblock {\em Proc. Natl. Acad. Sci.}, 107(16):7230--7234, 2010.

\bibitem{Sukhovich_2009}
A.~Sukhovich, B.~Merheb, K.~Muralidharan, J.~O. Vasseur, Y.~Pennec, P.~A.
  Deymier, and J.~H. Page.
\newblock Experimental and theoretical evidence for subwavelength imaging in
  phononic crystals.
\newblock {\em Phys. Rev. Lett.}, 102:154301, 2009.

\bibitem{Lemoult_2011}
Fabrice Lemoult, Mathias Fink, and Geoffroy Lerosey.
\newblock Acoustic resonators for far-field control of sound on a subwavelength
  scale.
\newblock {\em Phys. Rev. Lett.}, 107:064301, 2011.

\bibitem{Zhu_2011}
J.~Zhu, J.~Christensen, J.~Jung, L.~Martin-Moreno, X.~Yin, L.~Fok, X.~Zhang,
  and F.~J. Garcia-Vidal.
\newblock A holey-structured metamaterial for acoustic deep-subwavelength
  imaging.
\newblock {\em Nat. Phys.}, 7(1):52--55, 2011.

\bibitem{Hussein_2014}
M.~I. Hussein, M.~J. Leamy, and M.~Ruzzene.
\newblock Dynamics of phononic materials and structures: Historical origins,
  recent progress, and future outlook.
\newblock {\em Appl. Mech. Rev.}, 66(4):040802, 2014.

\bibitem{GdG_2011}
J.~M. Gómez-de Gabriel, A.~Mandow, J.~Fernández-Lozano, and
  A.~Garcia-Cerezo.
\newblock Using {LEGO NXT} mobile robots with {LabVIEW} for undergraduate
  courses on mechatronics.
\newblock {\em IEEE Trans. Edu.}, 54(1):41--47, 2011.

\bibitem{Gothelf_2012}
K.~V. Gothelf.
\newblock {LEGO}-like {DNA} structures.
\newblock {\em Science}, 338(6111):1159--1160, 2012.

\bibitem{Lind_2014}
K.~R. Lind, T.~Sizmur, S.~Benomar, A.~Miller, and L.~Cademartiri.
\newblock {LEGO}\textsuperscript{\sffamily\textregistered} bricks as building
  blocks for centimeter-scale biological environments: The case of plants.
\newblock {\em PLoS ONE}, 9(6):e100867, 2014.

\bibitem{Quercioli_1998}
F.~Quercioli, B.~Tiribilli, A.~Mannoni, and S.~Acciai.
\newblock Optomechanics with {LEGO}.
\newblock {\em Appl. Opt.}, 37(16):3408--3416, 1998.

\bibitem{Chen_2014}
B.~G. Chen, N.~Upadhyaya, and V.~Vitelli.
\newblock Nonlinear conduction via solitons in a topological mechanical
  insulator.
\newblock {\em Proc. Natl. Acad. Sci.}, 111(36):13004--13009, 2014.

\bibitem{Kokkonen_2007}
K.~Kokkonen, M.~Kaivola, S.~Benchabane, A.~Khelif, and V.~Laude.
\newblock Scattering of surface acoustic waves by a phononic crystal revealed
  by heterodyne interferometry.
\newblock {\em Appl. Phys. Lett.}, 91(8):083517, 2007.

\bibitem{Oudich_2011}
M.~Oudich, M.~Senesi, M.~B. Assouar, M.~Ruzenne, J.-H. Sun, B.~Vincent, Z.~Hou,
  and T.-T. Wu.
\newblock Experimental evidence of locally resonant sonic band gap in
  two-dimensional phononic stubbed plates.
\newblock {\em Phys. Rev. B}, 84:165136, 2011.

\bibitem{Assouar_2012}
M.~B. Assouar, M.~Senesi, M.~Oudich, M.~Ruzzene, and Z.~Hou.
\newblock {Broadband plate-type acoustic metamaterial for low-frequency sound
  attenuation}.
\newblock {\em Appl. Phys. Lett.}, 101(17):173505, 2012.

\bibitem{Casadei_2012}
F.~Casadei, T.~Delpero, A.~Bergamini, P.~Ermanni, and M.~Ruzzene.
\newblock Piezoelectric resonator arrays for tunable acoustic waveguides and
  metamaterials.
\newblock {\em J. Appl. Phys.}, 112(6):064902, 2012.

\bibitem{Celli_JSV_2014}
P.~Celli and S.~Gonella.
\newblock Laser-enabled experimental wavefield reconstruction in
  two-dimensional phononic crystals.
\newblock {\em J. Sound Vib.}, 333(1):114--123, 2014.

\bibitem{Rupin_2014}
M.~Rupin, F.~Lemoult, G.~Lerosey, and P.~Roux.
\newblock Experimental demonstration of ordered and disordered multiresonant
  metamaterials for lamb waves.
\newblock {\em Phys. Rev. Lett.}, 112:234301, 2014.

\bibitem{Pennec_2008}
Y.~Pennec, B.~Djafari-Rouhani, H.~Larabi, J.~O. Vasseur, and A.~C.
  Hladky-Hennion.
\newblock Low-frequency gaps in a phononic crystal constituted of cylindrical
  dots deposited on a thin homogeneous plate.
\newblock {\em Phys. Rev. B}, 78:104105, 2008.

\bibitem{Wu_2008}
Tsung-Tsong Wu, Zi-Gui Huang, Tzu-Chin Tsai, and Tzung-Chen Wu.
\newblock Evidence of complete band gap and resonances in a plate with periodic
  stubbed surface.
\newblock {\em Appl. Phys. Lett.}, 93(11):111902, 2008.

\bibitem{suppl}
See supplemental material (in tail of this document) for a more detailed
  account on the experimental setup and for additional results.

\bibitem{Torres_1999}
M.~Torres, F.~R. Montero~de Espinosa, D.~Garc\'ia-Pablos, and N.~Garc\'ia.
\newblock Sonic band gaps in finite elastic media: Surface states and
  localization phenomena in linear and point defects.
\newblock {\em Phys. Rev. Lett.}, 82:3054--3057, 1999.

\bibitem{Jensen_2003}
J.~S. Jensen.
\newblock Phononic band gaps and vibrations in one- and two-dimensional
  mass-spring structures.
\newblock {\em J. Sound Vib.}, 266(5):1053--1078, 2003.

\bibitem{Khelif_2004}
A.~Khelif, A.~Choujaa, S.~Benchabane, B.~Djafari-Rouhani, and V.~Laude.
\newblock Guiding and bending of acoustic waves in highly confined phononic
  crystal waveguides.
\newblock {\em Appl. Phys. Lett.}, 84(22):4400--4402, 2004.

\bibitem{Pennec_2009}
Y.~Pennec, B.~Djafari~Rouhani, H.~Larabi, A.~Akjouj, J.~N. Gillet, J.~O.
  Vasseur, and G.~Thabet.
\newblock Phonon transport and waveguiding in a phononic crystal made up of
  cylindrical dots on a thin homogeneous plate.
\newblock {\em Phys. Rev. B}, 80:144302, 2009.

\bibitem{Lemoult_2013}
F.~Lemoult, N.~Kaina, M.~Fink, and G.~Lerosey.
\newblock Wave propagation control at the deep subwavelength scale in
  metamaterials.
\newblock {\em Nat. Phys.}, 9(1):55--60, 2013.

\bibitem{Meseguer_1999}
F.~Meseguer, M.~Holgado, D.~Caballero, N.~Benaches, J.~S\'{a}nchez-Dehesa,
  C.~L\'{o}pez, and J.~Llinares.
\newblock Rayleigh-wave attenuation by a semi-infinite two-dimensional
  elastic-band-gap crystal.
\newblock {\em Phys. Rev. B}, 59:12169--12172, 1999.

\bibitem{Brule_2014}
S.~Br\^{u}l\'{e}, E.~H. Javelaud, S.~Enoch, and S.~Guenneau.
\newblock Experiments on seismic metamaterials: Molding surface waves.
\newblock {\em Phys. Rev. Lett.}, 112:133901, 2014.

\end{thebibliography}}

% SUPPLEMENTARY MATERIAL SECTION
\clearpage

% RESET FIGURE COUNTER TO 0
\setcounter{figure}{0}
\setcounter{page}{1}
\renewcommand{\thefigure}{S\arabic{figure}}
\renewcommand{\theequation}{S\arabic{equation}}
\renewcommand{\thepage}{S\arabic{page}}

\onecolumn
\section*{\Large Supplemental material (SM)}
\subsection*{Details on the experimental setup}
To facilitate the reproduction of our experimental results, we here report details regarding the data acquisition settings we chose in the Polytec PSV 9.0 Acquisition software. The acquisition is performed directly in the frequency domain (the Fast Fourier Transform is performed automatically within the acquisition system by the software). As already mentioned, to eliminate non-repeatable noisy features, measurements are repeated 20 times at each scanned location. We also resort to a low-pass filter with cutoff at $23\,\mathrm{kHz}$. As far as the vibrometer channel is concerned, we select a $5\,\mathrm{V}$ range and AC coupling. We acquire in the $0\mbox{--}20\,\mathrm{kHz}$ frequency range, and we concentrate on the $0\mbox{--}14\,\mathrm{kHz}$ band in postprocessing. The sampling frequency is $f_s=51.2\,\mathrm{kHz}$ and the number of FFT lines is 1600, resulting in a frequency resolution of $12.5\,\mathrm{Hz}$. The selected velocity decoder is the digital VD-08-$10\,\mathrm{mm/s/V}$, that allows acquisition up to $20\,\mathrm{kHz}$. The excitation is a pseudorandom waveform with maximum amplitude of $650\,\mathrm{mV}$. The excitation signal is amplified using a Bruel \& Kjaer Type $2718$ Power Amplifier, with gain set to $30\,\mathrm{dB}$.

\subsection*{Wave characterization of the baseplate}
\begin{figure} [!htb]
\centering
\includegraphics[scale=1.48]{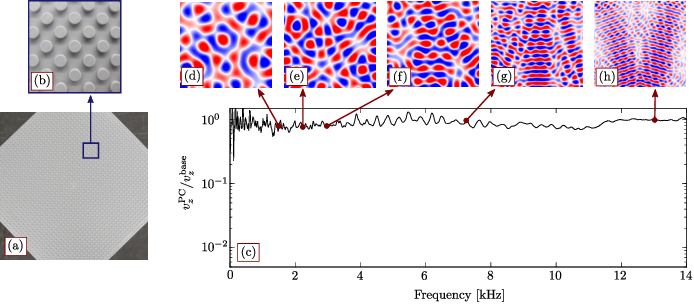}
\caption{Steady-state analysis of the brick-free baseplate. (a,b) Baseplate and detail of the studs. (c) Frequency response function (average out-of-plane velocity normalized by the average value measured along the bottom edge). (d-h) Laser-acquired wavefields sampled at the same five reference frequencies analyzed in Figs.~2d$\mbox{--}$h. (d) $1.5\,\mathrm{kHz}$. (e) $2.25\,\mathrm{kHz}$. (f) $3\,\mathrm{kHz}$. (g) $7.25\,\mathrm{kHz}$. (h) $13\,\mathrm{kHz}$.}
\label{SFig1}
\end{figure}

To further confirm the brick-dependent nature of the bandgaps discussed throughout this work, we analyze the response of the baseplate with no bricks attached. The plate and a detail of the studs arrangement is shown in Figs.~\ref{SFig1}a and b.
In Fig.~\ref{SFig1}c we show the frequency response function of the baseplate (for the FRF we use the same output metric used for the other topologies across the manuscript). Once we remove the bricks, we clearly see that no bandgaps are present in the $0\mbox{--}14\,\mathrm{kHz}$ range. This aspect is corroborated by the snapshots of the plate's steady-state response shown in Figs.~\ref{SFig1}d$\mbox{--}$h: the structure is able to propagate waves at all frequencies, including those that, in the \emph{dense} topology, corresponded to bandgap regions (Figs.~\ref{SFig1}e and g, corresponding to $2.25\,\mathrm{kHz}$ and $7.25\,\mathrm{kHz}$, respectively). Note that, due to the presence of the small cylindrical studs, the baseplate itself is a periodic structure; however, we did not detect any stud-related gap in the considered frequency range, which suggests that the amplitude of the associated scattering events is negligible.

\begin{figure} [!htb]
\centering
\includegraphics[scale=1.48]{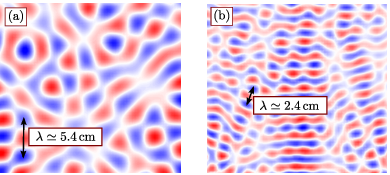}
\caption{Wavelength identification from the baseplate response. (a) Baseplate response at $2.25\,\mathrm{kHz}$, inside the locally resonant bandgap of the dense topology. (a) Baseplate response at $7.25\,\mathrm{kHz}$, inside the Bragg bandgap of the dense topology.}
\label{SFig2}
\end{figure}

We now use the response of the pristine plate to visually quantify the wavelength of plate waves having frequencies that fall inside bandgaps observed in the dense topology. Note that, for the dense topology, the unit cell characteristic dimension is $a=1.2\,\mathrm{cm}$. In Fig.~\ref{SFig2}a we zoom on the baseplate response at $2.25\,\mathrm{kHz}$, corresponding to the locally resonant gap. 
Note that in Fig.~\ref{SFig2}, for visualization purposes, the velocity amplitude bar has been modified to range from $-v^{\mathrm{base}}_{z,\,\mathrm{max}}$ to $v^{\mathrm{base}}_{z,\,\mathrm{max}}$. From the drawn arrow, we can see that the characteristic wavelength of the response can be estimated to be $\lambda\simeq5.4\,\mathrm{cm}$, a value corresponding to about $6$ unit cells. In Fig.~\ref{SFig2}a we report the baseplate response at $7.25\,\mathrm{kHz}$, corresponding to the Bragg gap. In this case, the estimated wavelength is $\lambda\simeq2.4\,\mathrm{cm}$, corresponding to $2$ unit cells. This result appears to support the notion that locally resonant gaps are subwavelength in nature, while Bragg ones take place when the wavelength is comparable to the unit cell size.

\begin{figure} [htb]
\centering
\includegraphics[scale=0.74]{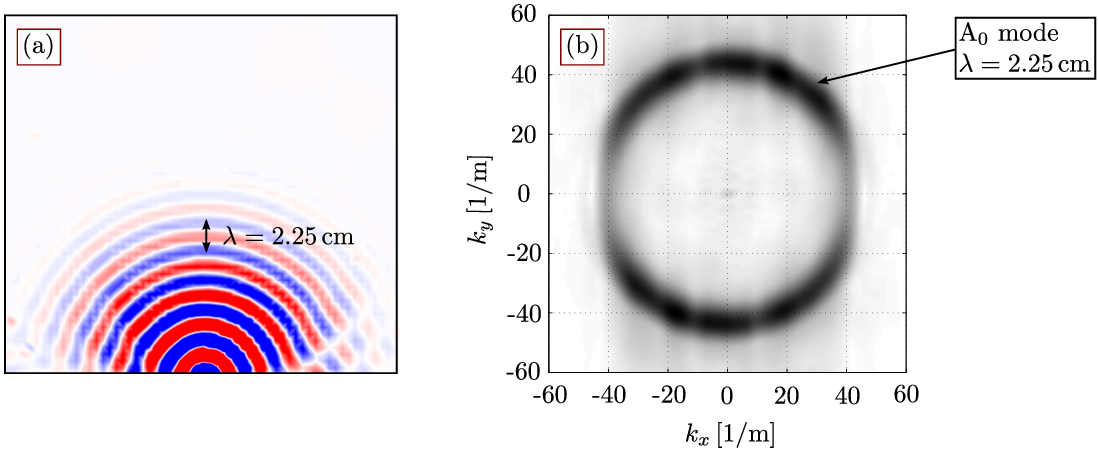}
\caption{Transient analysis of the pristine baseplate in response to a 5-cycle burst with carrier frequency $6.75\,\mathrm{kHz}$. (a) Wavefield at $t=0.70\,\mathrm{ms}$. (b) 2D-FFT of the wavefield in (a).}
\label{f}
\end{figure}

It is also interesting to briefly discuss the modal characteristics of the wave response of the pristine baseplate. In theory, we expect two modes to coexist in the low-frequency interval considered in our experiments: the first antisymmetric Lamb mode, $\mathrm{A_0}$, and the first symmetric Lamb mode, $\mathrm{S_0}$. However, given the specific excitation configuration (the direction of actuation is here perpendicular to the plate), the response is dominated by the antisymmetric mode [Rupin et al, Phys. Rev. Lett. 112, 234301, 2014]. In fact, the $\mathrm{A_0}$ characteristics at low frequencies are those of flexural waves, that can be alternatively captured by plate theory models (e.g. the Kirchhoff model), which in fact entail a single mode corresponding to the out-of-plane displacement of points on the midplane of the plate. To highlight the mono-modal nature of the plate response, we report in Fig.~\ref{f}a a \emph{transient} wavefield obtained in the pristine baseplate excited with a 5-cycle burst with carrier frequency $6.75\,\mathrm{kHz}$.
This frequency is selected in the middle of the range used in our analysis and can be safely considered representative of the plate behavior in the entire low-frequency domain. Note that the experimental setup is slightly different from the setup used throughout this letter, in terms of specimen clamping, location of excitation and scan grid selection. To investigate the modes of wave propagation, we compute the 2D-FFT of the wavefield, shown in Fig.~\ref{f}b. We can see that the spectral amplitude in the wavenumber space manifests indeed as a largely single-modal feature, i.e. a ring with radius corresponding to the reciprocal of the wavelength observed in Fig.~\ref{f}a.

\subsection*{Effect of the resonators' height on the bandgap behavior}
\begin{figure} [htb]
\centering
\includegraphics[scale=1.48]{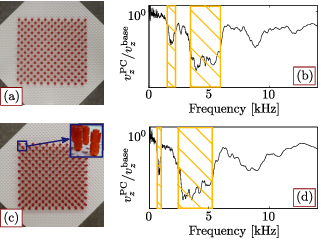}
\caption{Effects of the resonators' height. (a) Coarse periodic topology with single-brick resonators. (b) FRF for the topology in (a). (c) Coarse periodic topology with double-brick resonators; a detail of the brick stacking is also shown. (d) FRF for the topology in (c).}
\label{SFig3}
\end{figure}

We here investigate the effect of the resonators' height on the bandgap behavior of the coarse periodic topology shown in Fig.~3c. In particular, we compare the ``single-brick'' case, shown again in Fig.~\ref{SFig3}a, to a ``double-brick'' case in which all resonators are made by stacking two bricks on top of each other, as shown in Fig.~\ref{SFig3}c.
Figs.~\ref{SFig3}b and d represent the FRFs for the single- and double-brick cases, respectively. A direct comparison of the two plots highlights how the locally resonant bandgap shifts towards lower frequencies as the resonator height increases. This phenomenon can be explained in light of the drop in resonance frequency of the pillar, resulting from the increase in mass moment of inertia due to the doubling of the mass and length of the resonator. This further confirms the locally resonant nature of the lowest gap and hints at a rigid tilting motion of the pillars as the dominant resonant mechanism. It is worth noting that the Bragg bandgap also shifts slightly towards lower frequencies as a consequence of the change in the impedance mismatch between baseplate and pillars (due to the change in inertial properties of the latter).

\subsection*{Crystal topologies with point defects}
\begin{figure} [htb]
\centering
\includegraphics[scale=1.48]{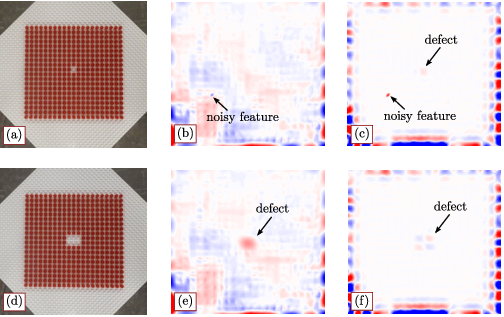}
\caption{Experimental evidence of defect modes in phononic crystals with point defects. (a) Topology with small defect, (b) its response at $2512.5\,\mathrm{Hz}$ and (c) its response at $7000\,\mathrm{Hz}$. (d) Topology with large defect, (e) its response at $2512.5\,\mathrm{Hz}$ and (f) its response at $7000\,\mathrm{Hz}$.}
\label{SFig4}
\end{figure}

We further expand the portfolio of architectures that we can study with this approach by considering two additional topologies, both obtained by introducing point defects of various sizes within a dense topology. The specimen in Fig.~\ref{SFig4}a is obtained by removing a single brick. 
Its response in the locally resonant bandgap region (Fig.~\ref{SFig4}b) shows that the defect is not large enough to resolve the energy localization at the corresponding wavelength. In the case of Fig.~\ref{SFig4}c, corresponding to the Bragg gap, we can see how the energy localizes at the defect site. Note that, in this set of results, we can see a spurious noisy feature in the wavefield data. The persistence of this feature across the frequency spectrum indicates that this is an artifact of the acquisition process, probably due to a failed laser measurement at one of the scan points, and not the signature of any abnormal behavior of the structure. The topology in Fig.~\ref{SFig4}d features a $3\times2$ defect, obtained by removing 6 bricks. In Figs.~\ref{SFig4}e and f, we can see that the defect is large enough to induce energy localization in both locally resonant and Bragg bandgap regimes.

\end{document}